\documentclass[twoside]{dis04}
\usepackage{amsmath}
\def\runauthor{Andre Georg Holzner}
\def\shorttitle{Towards Precision Measurements at the LHC}

\def\nobs{\ensuremath{N_\mathrm{observed}}}

\def\nbg{\ensuremath{N_\mathrm{background}}}

\def\nsigexp{\ensuremath{N_\mathrm{signal}^\mathrm{expected}}}
\def\nsigobs{\ensuremath{N_\mathrm{signal}^\mathrm{observed}}}
\def\nsigcorr{\ensuremath{N_\mathrm{signal}^\mathrm{corrected}}}

\newcommand{\eff}[1]{\ensuremath{\varepsilon_\mathrm{#1}}}
\def\Lpp{\ensuremath{L_{\mathrm{pp}}}}
\newcommand{\xsect}[1]{\ensuremath{\sigma_\mathrm{#1}}}
\begin{document}

\title{Towards Precision Measurements at the LHC}

\author{ANDRE~GEORG~HOLZNER}

\address{
  Institut f\"ur Teilchenphysik,
  ETH H\"onggerberg\\
  CH-8093 Z\"urich, Switzerland\\
  E-mail: Andre.Georg.Holzner@cern.ch\\
    }

\maketitle

\abstracts{This article discusses some basic aspects of cross
  section measurements at the future Large Hadron Collider (LHC) at CERN. }


\section{Cross section measurements}
  One basic physics problem in collider experiments is 
  to measure cross sections, thus to count the number of observed events
  after a given selection. The selected event sample is not 
  background-free in most cases, so one has to perform some kind of background
  subtraction:
  \begin{equation}
    \nsigcorr = \dfrac{\nsigobs}{\eff{signal}} = \dfrac{\nobs - \nbg}{\eff{signal}}
  \end{equation}

  Where \nsigcorr{} is the estimate of the number of events 
  coming from the process of interest ('signal') produced in the detector.
  \nsigobs{} is the number of `observed signal events', \nbg{} is the number of background events (expected or measured)
  and \eff{signal} is the
  efficiency of the overall selection (trigger and off-line
  selection efficiency, detector acceptance etc.) for the signal process. Usually,
  one wants to compare \nsigobs{} to the expected number \nsigexp{}:

  \begin{equation}
    \nsigexp = \xsect{partons\to signal} \otimes
    PDF(x_1,x_2,Q^2) \times \Lpp{} \times \eff{signal}
  \end{equation}

  where and $\otimes$ is in fact a convolution of the parton
  distribution functions (PDFs) 
  and the hard cross section. \Lpp{} is the proton-proton luminosity,
  which can be measured as:

  \begin{equation}
    \Lpp{} = \dfrac{N_\mathrm{pp\to pp (+ X)}}
      {\eff{pp\to pp (+ X)} \xsect{pp\to pp (+ X)}}
  \end{equation}

  i.e.{} by counting the number of (quasi) elastic 
  proton-proton collisions.

  The problem with the proton-proton luminosity however is that
  it is difficult to measure and to calculate accurately. In fact, the ATLAS and CMS
  collaborations, currently estimate to achieve about 5\% uncertainty
  on \Lpp{}~\cite{lumi-tdrs}.
  This would
  mean that absolute cross sections can not be measured with an accuracy
  better than 5\%.


\section{Luminosity measurements}

  The question arises whether this limitation of the proton-proton
  luminosity is actually important. In fact, considering that
  for all calculations (free) partons collide with each other at LHC energies
  (as opposed to protons), one should be looking for a 
  precise determination of the {\sl parton luminosity}~\cite{Dittmar:1997md}.
  In other words, use a `hard' process (instead of 
  quasi elastic proton-proton scattering) to normalize 
  the cross sections to. Such a process must fulfill the
  following conditions:

  \begin{enumerate}
  \item It must have a high rate
  \item It must have a clean signature with small background, and last
    but not least,
  \item Precise calculations for the (differential) cross section must
    exist.
  \end{enumerate}


  Single W and Z production clearly fulfill condition 1.{}  and 2.{} 
and their couplings to fermions have been measured at LEP 
to an accuracy of 1\% or better.
 More precise calculations are becoming available~\cite{Anastasiou:2003ds}.

  Let us consider an example: The number of W pair and single
  Z events is given by:

  \begin{equation}
    \begin{array}{llllllllll}
    N_\mathrm{pp\to WW}^\mathrm{expected} & = &  & \sigma_\mathrm{pp\to WW} & \otimes & PDF(x_1',x_2',Q'^2) & \times & \Lpp{} & \times & \eff{pp\to WW} \\ 
   \\
    N_\mathrm{pp\to Z}^\mathrm{observed}  & = &  & \sigma_\mathrm{pp\to Z}  & \otimes & PDF(x_1,x_2,Q^2)    & \times & \Lpp{} & \times & \eff{pp\to Z}  \\
    \end{array}
\end{equation}

  Dividing the first equation by the second, one obtains:
  
  \begin{equation}
    N_\mathrm{pp\to WW}^\mathrm{expected} = 
    N_\mathrm{pp\to Z}^\mathrm{observed} 
    \times 
    \dfrac{\sigma_\mathrm{pp\to WW}}{\sigma_\mathrm{pp\to Z}}
    \times 
    \dfrac{\eff{pp\to WW}}{\eff{pp\to Z}}
    \otimes
    \dfrac{PDF(x_1',x_2',Q'^2)}{PDF(x_1,x_2,Q^2)}
  \end{equation}

  i.e. the number of expected W pair events is expressed 
  in terms of the number of observed single Z events, 
  the cross section ratios and a PDF ratio. The proton-proton
  luminosity cancels out. The systematic uncertainties which are
  left come from the selection efficiencies, the theoretical cross 
  section predictions and the PDF uncertainties. 


\section{PDF uncertainties}

\begin{figure}[thb]
\vspace*{3mm}
\begin{center}
\centerline{
 \epsfig{width=0.4\linewidth,file=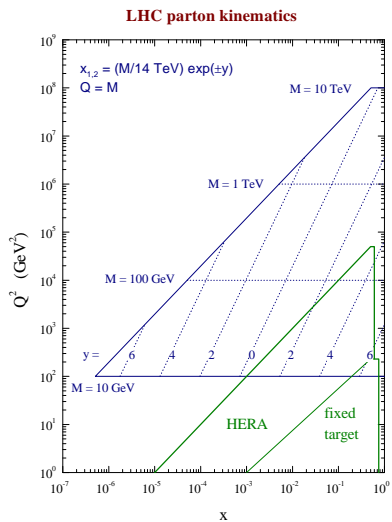}
}
\caption[*]{Relation between rapidity $y$, the scale of the 
  hard interaction $Q$ (here set to the mass) and the momentum fraction of the
  initial-state parton in heavy particle production at
  LHC~\cite{Thorne:2003gu}. 
  The graph shows the 
  regions measured by the HERA experiments and the region important
  at LHC.
\label{fig:lhc-kinematics}
}
\end{center}
\end{figure}

Fig.~\ref{fig:lhc-kinematics} shows the regions in the 
$x$ vs. $Q^2$ plane covered by past and present experiments.
It can be seen clearly that a large fraction of the region
accessible at LHC is uncovered by today's experiments and 
thus one has to rely on extrapolations of today's experiments
(at lower $Q^2$) to LHC scales. For example, to produce a W boson
at rapidity 0, both partons have a $x$ of 0.006. If one goes
to non-zero rapidities, one of the partons must have a smaller $x$.

The uncertainties are significantly larger for $x < 0.005$ 
(e.g.\ in the MRST PDFs) than for $x > 0.005$. This due to inconsistencies in the data points
fitted, which affects a wide kinematic region of interest
at the LHC. These uncertainties are expected to be reduced by 
including higher order (full NNLO) calculations, theoretical
corrections for extremely small and large $x$ as well
as corrections at low $Q^2$~\cite{Martin:2003sk}.

One approach to estimate the consequences of such uncertainties
at LHC scales is to calculate cross sections with different
PDFs and compare the values obtained. Since recently, several PDF functions provided by the fitting groups 
now also include an uncertainty (e.g.{}~\cite{Martin:2002aw}).


\section{Constraining PDFs at LHC}

When LHC becomes operational, the PDFs can be constrained further 
from the data itself instead of solely relying on the extrapolations
based on today's measurements.

\subsection{Quarks}
  
\begin{figure}[thb]
\begin{center}
\centerline{
 \epsfig{width=0.5\linewidth,file=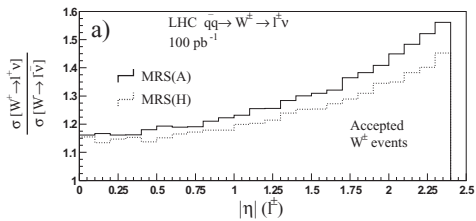}
}
\caption[*]{Ratio $\sigma({\mathrm{pp\to W^+ \to \ell^+ \nu}}) /
    \sigma(\mathrm{pp\to W^- \to \ell^- \bar{\nu}})$ as function 
    of the charged lepton pseudorapidity for two very similar
    PDFs~\cite{Dittmar:1997md}.
\label{fig:single-w-ratio-rapi}
}
\end{center}
\end{figure}

Single W and Z production are perfect processes to constrain
the relative quark densities at the LHC (e.g. the ratio of the up- to the down-quark density). 
For a fixed interaction scale $Q$ (i.e.{} particle mass), the product $x_1x_2$ of the two momentum fractions
of the colliding partons is fixed (in leading order). The rapidity of the particle 
is then determined by the ratio $x_1 / x_2$ (see Fig.~\ref{fig:lhc-kinematics}). 
Thus, different rapidities of heavy particles are sensitive to different ranges of $x$
values. An example of the ratio of the number of charged leptons produced 
in single $\mathrm{W}^\pm$ events in different pseudorapidity bins
of the charged lepton is shown in Fig.~\ref{fig:single-w-ratio-rapi}.
The PDFs shown differ only in their sea quark distributions which 
is either symmetric (MRS(A)) or not (MRS(H)).
Although they are not the latest sets of PDFs, they nicely illustrate
that only a very small amount of data is needed to distinguish between very similar PDFs.
The statistical uncertainty is about 1\% in each bin for data corresponding
to roughly one day in the  low-luminosity phase of LHC. 


\subsection{Gluons}

\begin{figure}[thb]
\begin{center}
\centerline{
 \epsfig{width=0.5\linewidth,file=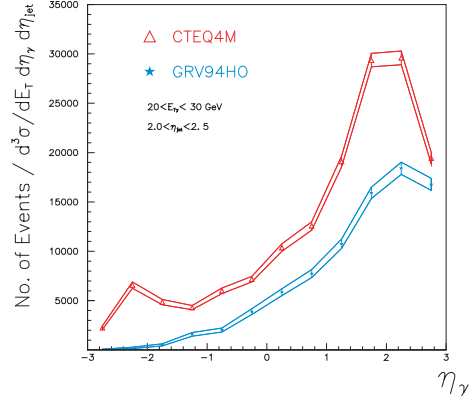}
}
\caption[*]{Photon pseudorapidity for a photon + jets
selection for two different PDFs~\cite{heath-gluon-pdfs}.
The number of events shown correspond to 10 days of
luminosity of $10^{32}\mathrm{cm}^{-2} \mathrm{s}^{-1}$.
\label{fig:gluon-pdf-variable}
}
\end{center}
\end{figure}

About half of the proton's momentum is carried by
gluons. Furthermore, the gluon distributions are
often determined only indirectly in deep inelastic
scattering experiments. It is therefore important
to determine the gluon distribution directly at the LHC.
This can be done using processes like $\mathrm{g + q} \to
\mathrm{q + (\gamma/W/Z)}$
which correspond to the experimental signature  
jet + photon / W / Z. For example photons can be measured
very precisely with the ATLAS and CMS detectors. 
Fig.~\ref{fig:gluon-pdf-variable} shows the photon
pseudorapidity distribution after a cut on the 
photon energy and the jet pseudorapidity
for two different PDFs~\cite{heath-gluon-pdfs}. 
The main background are events with a leading $\pi^0$ looking
like single isolated photon in the detector. 
The absolute scale has an uncertainty of about 10\% 
which comes from the choice of the QCD renormalization scale.

\section{Higher Order Calculations}

\begin{figure}[!thb]
\begin{center}
\centerline{
 \epsfig{width=0.5\linewidth,file=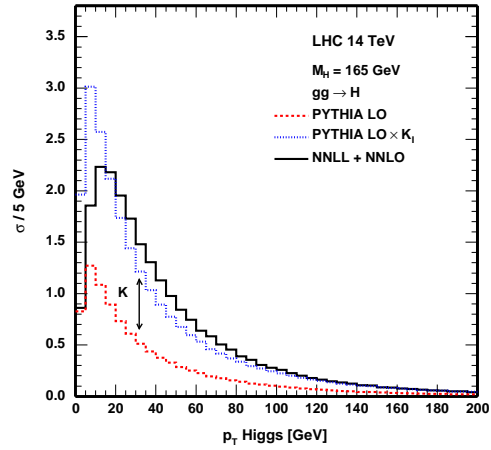}
}
\caption[*]{This plot illustrates the
   need for a differential higher order cross section calculation~\cite{Davatz:2004zg}:
   The leading order differential cross section
   scaled by the K factor does not reproduce the NNLL + NNLO
   calculation. Thresholds on the transverse energies of the trigger 
   will preferably select high transverse momentum Higgs 
   events, whose fraction is underestimated by the 
   LO $\times$ K-factor distribution.
  \label{fig:higgs-pt}
}
\end{center}
\end{figure}

Most of the parton shower Monte Carlo generators available today
are based on leading order (LO) calculations.
Higher order calculations are often available for the
total cross section. A widely used procedure to study how well 
a certain physics process can be observed or measured at LHC
is to use a leading order parton shower Monte Carlo generator
and scale the distributions of the observables such that
the cross section calculated by the generator matches analytical
higher order results (the scaling factor is commonly known as
'K-Factor'). In practice, one is however obliged to apply some
cuts on the transverse momentum (e.g. due to the trigger threshold) 
or the pseudorapidity of the particles produced (due to limited
detector coverage). An example of a differential cross section
obtained with PYTHIA~\cite{Sjostrand:2000wi} and from analytical higher order
calculations is shown in Fig.~\ref{fig:higgs-pt}. The simply scaled LO 
distribution does not match the higher order analytical calculation
and thus introduces a large uncertainty of the selection efficiency. However,
a reweighting method allows perfect matching.

\section{Summary}

Todays uncertainties of PDFs
are about 4\%~\cite{Martin:2003sk}.
Uncertainties on single W and Z cross sections 
due to the experimental uncertainty in the PDFs amount 
to 3\%, 
ratio measurements can be better, e.g. 0.5\%. 

While today we can only extrapolate the PDFs from measurements
performed e.g. by HERA to the LHC scale, the latter will be 
able to constrain them further in the kinematic region of interest.
Single W and Z production are a useful tool to constrain the quark
distributions while jet + photon events give a handle on the 
gluon distribution.

An optimistic estimate of the experimental precision achievable
for single W and Z measurements is 1\%, which can be used reduce the 
uncertainty on the recorded luminosity to values significantly smaller
than the 5\% previously foreseen.

NNLO calculations will be necessary wherever
a quantity can be measured to better than 10\% accuracy
experimentally if the theoretical error should be of the
order of the experimental error.


\end{document}